# Programmable Beam Control for Electron Energy-Loss Spectroscopy and Ptychography

Mariana Palos,[1] Liam Spillane,[2] Geri Topore,[1] Yaqi Li,[1] David Pesquera,[5] Colin Ophus,[3] Stephanie M. Ribet[4] and Michele Shelly Conroy[1],*

[1]Department of Materials, London Centre of Nanotechnology, Imperial Henry Royce Institute, Imperial College London, London, SW7 2AZ, United Kingdom, [2]Gatan Inc., Pleasanton, CA, USA, [3]Department of Materials Science and Engineering, Stanford University, Stanford, CA, 94305, USA, [4]National Center for Electron Microscopy, Molecular Foundry, Lawrence Berkeley National Laboratory, Berkeley, CA, 94720, USA and [5]Catalan Institute of Nanoscience and Nanotechnology, ICN2, 08193, Bellaterra, Barcelona ,Spain

*Corresponding author. mconroy@imperial.ac.uk

**Abstract**

Programmable electron-beam scanning offers new opportunities to improve dose efficiency and suppress scan-induced artifacts in scanning transmission electron microscopy. Here, we systematically benchmark the impact of non-raster trajectories, including spiral and multi-pass sequential patterns, on two dose-sensitive techniques: electron energy-loss spectroscopy (EELS) and ptychography. Using $DyScO_3$ as a model perovskite, we compare spatial resolution, spectral fidelity, and artifact suppression across scan modes. Ptychographic phase reconstructions consistently achieve atomic resolution and remain robust to large jumps in probe position. In contrast, atomic-resolution EELS maps show pronounced sensitivity to probe motion, with sequential and spiral scans introducing non-uniform elemental contrast. Finally, spiral scanning applied under cryogenic conditions in $BaTiO_3$ thin films improves dose uniformity and mitigates drift-related distortions. These results establish practical guidelines for the implementation of programmable scan strategies in low-dose 4D-STEM and highlight the inherent resilience of ptychography to trajectory-induced artifacts.

**Key words:** Scanning transmission electron microscopy (STEM), 4D-STEM, programmable beam scanning, electron energy-loss spectroscopy (EELS), ptychography, radiation damage, low-dose imaging, cryogenic microscopy

## Introduction

Electron energy-loss spectroscopy (EELS) and four-dimensional scanning transmission electron microscopy (4D-STEM) are complementary, yet dose-sensitive techniques in modern electron microscopy. EELS provides insight into electronic structure, bonding, and low-energy excitations such as phonons. Monochromated systems can achieve meV energy resolution and enable atomic-scale chemical mapping through advanced quantification methods [1, 2, 3]. 4D-STEM records a diffraction pattern at each probe position, providing access to position-sensitive structural information, which can be used to map ordering in materials at high-resolutions including probing strain and electromagnetic field information. 4D-STEM-based phase retrieval methods, such as iterative electron ptychography, allow for high-resolution, dose-efficient characterization of structure through reconstruction of real space potential [4, 5, 6, 7].

Both EELS and 4D-STEM demand relatively high electron doses, making beam-induced damage a critical limitation, particularly for organic, hybrid, and oxide materials where radiolysis dominates [8]. While low-dose workflows combining EELS and 4D-STEM are feasible, the spatial and temporal distribution of the dose can be as important as the total dose itself. Verbeeck et al. demonstrated that interleaved, spiral, and random scan trajectories can delay the onset of damage, reducing structural degradation in zeolites by up to 11% under equivalent total dose conditions [9, 10]. These findings underscore that controlling how, not just how much, dose is delivered can preserve structure and chemistry in beam-sensitive systems.

Cryogenic STEM provides a route for imaging fragile materials and stabilizing low-temperature phases without inducing structural changes. However, cryogenic workflows introduce additional challenges: increased drift, reduced signal-to-noise ratio (SNR), and higher contamination risk. Under these conditions, precise alignment, active drift correction, and dose-efficient acquisition strategies become essential. Programmable scan paths, such as spiral and sparse, can mitigate these challenges by reducing localized hot spots and eliminating dwell time accumulation along raster flyback lines [11, 12, 13].

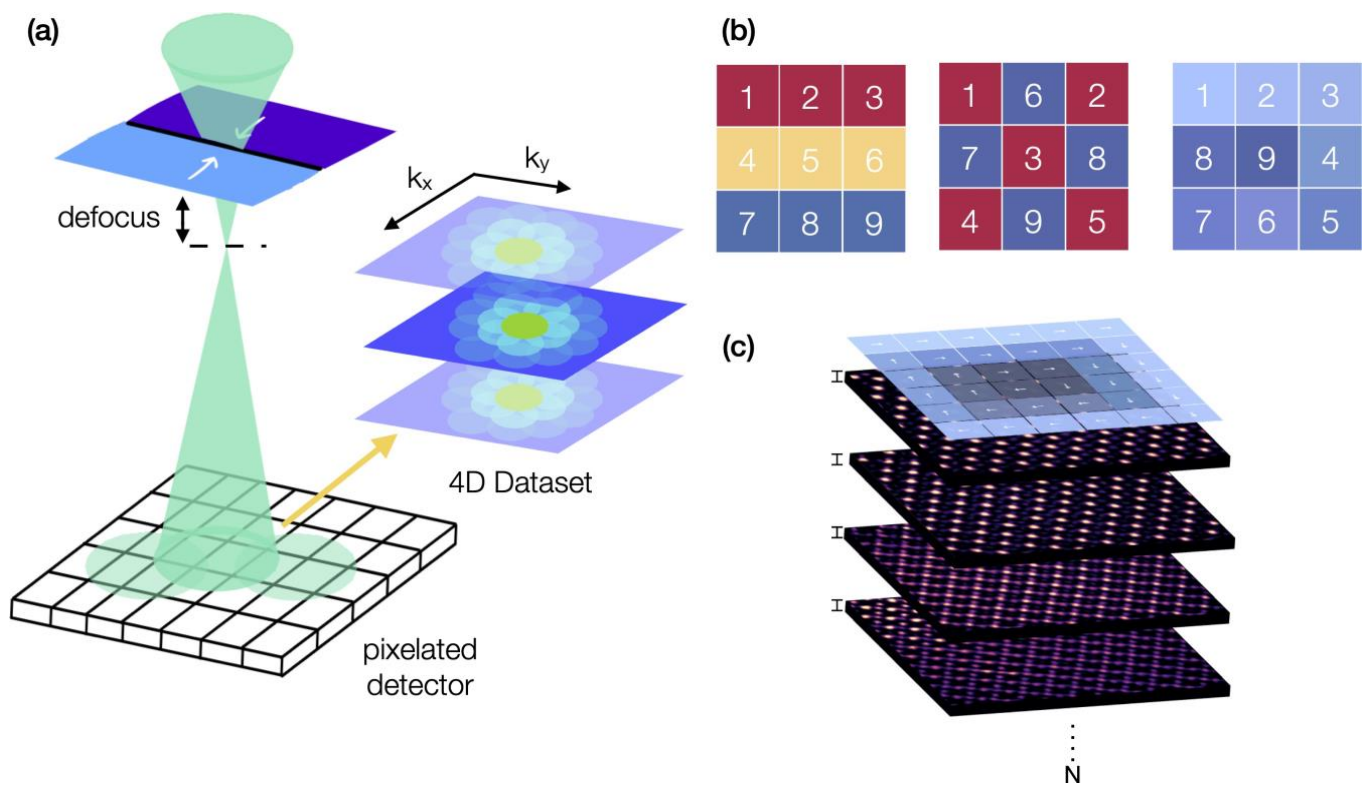

Fig. 1: (a) Schematic of the 4D-STEM experimental setup used in this study. A defocused electron probe is scanned across the sample, while 2D diffraction patterns are recorded at each probe position using a pixelated detector (Gatan K3) (b) Examples of programmable scan trajectories include raster, spiral (inward/outward), and N-sequential scan patterns. (c) Illustration of a multislice ptychography reconstruction, with a spiral scan path overlaid on the first slice, demonstrating compatibility between custom scan strategies and ptychographic phase retrieval.

The trajectory of the electron probe in STEM has a strong influence on the fidelity of both structural and spectroscopic data. Conventional raster scanning is prone to flyback delays, probe lag, and drift-induced distortions. To overcome these limitations, programmable scan patterns, such as spiral, N-sequential, sparse, random, and interleaved paths, have been developed to improve dose uniformity, reduce motion-induced distortion, and eliminate flyback artifacts [14, 15, 16, 17, 18, 10]. Implementing such trajectories requires precise synchronization of scan control, beam blankers, and high-speed detectors. Recent advances, including electromagnetic beam blankers [19], adaptive scanning [20, 21], flyback compensation algorithms [22], and event-driven acquisition with Timepix3 detectors [23] have made real-time and dose-efficient scanning increasingly practical.

Beyond imaging, tailored beam trajectories can play an important role in studies that explore how beam-specimen interactions actively modify materials, such as enabling manipulation of dopants and defects in 2D materials [24, 25, 26] and control of domain wall motion in ferroelectrics [27, 28, 29, 30, 31, 32]. Understanding how scan patterns influence both measurement quality and material response is therefore critical for designing robust experiments.

In this work, we evaluate how programmable scan patterns, including raster, inward/outward spiral, and N-sequential, affect spatial and spectral fidelity at low doses, with a focus on ptychographic phase retrieval and EELS (Figure 1). Using $DyScO_3$ at room temperature, we benchmark these strategies for spatial resolution and artifact suppression. We then applied spiral scanning to free-standing cryogenic $BaTiO_3$ thin films to assess their ability to resolve light elements while minimizing beam-induced damage. These results provide practical guidelines for optimizing scan trajectories in high-resolution, dose-sensitive STEM workflows.

## Experimental Results

### Atomic-Resolution EELS Mapping in $DyScO_3$

$DyScO_3$ was chosen as a model perovskite oxide system to benchmark programmable scan performance due to its well-defined crystal structure, the presence of relevant oxygen columns for quantification, and its stability under electron irradiation. Using raster, spiral and N-sequential scan patterns from Figure 2(a-f), we acquired atomic-resolution Sc L-edge and Dy M-edge EELS maps (Figure 3(g-l)). All scan trajectories produced interpretable elemental contrast. However, variations in signal continuity and intensity uniformity were evident. In several cases, discontinuities between adjacent atomic columns correlated with the underlying scan path. Additional elemental maps including oxygen are shown in Supplementary Figure S1.

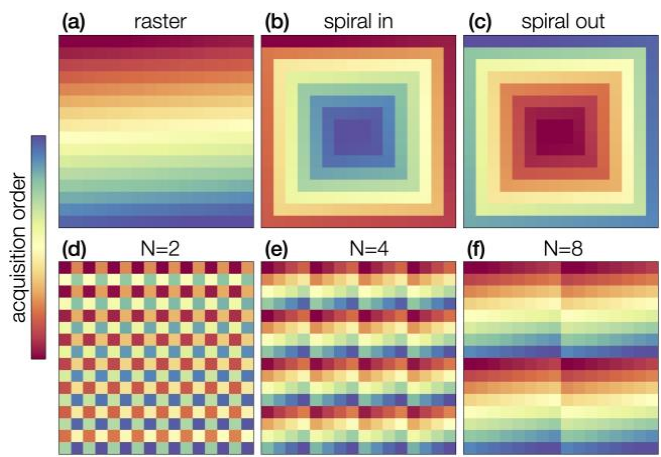

Fig. 2: Schematic of scan path variations: (a) raster, (b) inward spiral, (c) outward spiral, (d) 2×2 sequential, (e) 4×4 sequential, and (f) 8×8 sequential. The color scale indicates the acquisition order, progressing from red (first pixels acquired) to blue (last pixels acquired).

To further characterize these differences, fast Fourier transforms (FFTs) were computed (Figure 3(m-r)) from the simultaneously acquired ADF frames (Figure 3(a-f)). The FFTs revealed variations in spatial frequency completeness and symmetry. For example, in N = 8 sequential scans, where the ADF images exhibited pronounced discontinuities, the FFTs contained additional reflections resembling replicated Bragg peaks (Figure 3(o)). These likely arise from probe-settling delays during large pixel jumps, leading to positional inconsistencies that manifest as shifted replicas of true lattice reflections.

Comparing spiral and raster scans, raster trajectories retained faint vertical streaks in the FFTs even after filtering, consistent with line-by-line flyback hysteresis (red arrows in Figure 3(m)). The sharper, cleaner peaks in the spiral FFTs shown in Figure 3(r) confirm reduced sensitivity of spiral trajectories to coil hysteresis and settling-time distortions. These findings are consistent with previous reports that attribute such artifacts to coil hysteresis and probe settling delays during larger pixel jumps [14, 10, 17]. Such suppression of spatial-frequency artifacts induced by scan is critical, as these errors propagate directly into spectroscopic maps and ptychographic reconstructions, degrading both spatial and spectral fidelity.

These observations underscore the importance of evaluating scan pattern performance not only in real-space imaging but also in reciprocal space, where subtle distortions can compromise quantitative analysis. While advanced correction algorithms exist for both real-space and spectroscopic datasets, including geometric distortion correction, drift compensation, and denoising, [33, 34, 35, 36] here we assess whether ptychographic reconstruction alone can recover high-resolution information from alternative probe trajectories.

### Ptychographic reconstructions in DyScO$_3$

Ptychographic phase reconstructions from DyScO$_3$ (Figure 4) revealed minimal differences in achievable resolution across N-sequential scan patterns. Bragg peaks remained visible to ~0.7 Å even for N=8 sequential scans. Unlike ADF and EELS maps, ptychographic reconstructions largely suppressed trajectory-related distortions, reflecting their reliance on redundant, overlapping diffraction data rather than precise real-space probe positioning.

These findings underscore a critical distinction: while direct imaging modes preserve positional errors as visible artifacts, iterative ptychographic solvers can absorb moderate scan-path irregularities without compromising spatial resolution. Provided that probe overlap and reciprocal space redundancy remain sufficient, ptychography can recover high-resolution phase information even from "non-ideal" scan trajectories.

### Cryogenic Spiral Scanning in Free-standing BaTiO$_3$

To evaluate programmable scanning under beam-sensitive conditions, we next applied spiral scanning to free-standing BaTiO$_3$ thin films, imaged at cryogenic temperatures. Outward spiral scans produced ptychographic reconstructions that clearly resolved oxygen columns (Figure 5(c)), with minimal drift or distortion despite the increased mechanical and electrostatic instability typical of cryo-STEM operation. These reconstructions also revealed subtle oxygen and titanium displacements relative to their centrosymmetric cubic positions, consistent with the onset of a low-temperature ferroelectric phase, demonstrating how programmable scanning can recover delicate structural signatures in dose-sensitive oxide systems.

## Discussion

### Radiation-Damage Mechanisms and Temporal Dose Distribution

Electron-beam-induced damage in the microscope arises primarily from two mechanisms: (i) inelastic ionization (radiolysis) and (ii) elastic displacement (knock-on). Their relative importance depends on both the material chemistry and the acceleration voltage [37]. In programmable scanning, the temporal and spatial distribution of the dose is as important as the integrated dose, as it governs how local chemical, structural, and thermal responses accumulate.

Radiolytic processes are inherently time-dependent. Damage channels such as chemical bond breaking and desorption of volatile species can partially relax or redistribute between electron hits if illumination is spread out in time. In contrast, delivering many electrons to the same region in rapid succession increases the probability that those intermediate species react before they can diffuse away or recombine, thereby accelerating net damage [8]. In practice, scan trajectories where electron exposure is concentrated locally over short time intervals (e.g., looping repeatedly over a small region or effectively "parking" the probe) are more damaging than scans that spread the electron fluence across a larger area before returning. Scan trajectories where the electron exposure is distributed more evenly in time across neighboring pixels reduce the likelihood of instantaneous dose accumulation at any one site and delay the onset of irreversible changes. This principle underpins the advantage of non-raster scan strategies over conventional line-by-line acquisition for beam-sensitive systems [9, 10].

In the present work, DyScO$_3$ and BaTiO$_3$ are relatively robust oxide perovskites compared with organic or hybrid beam-sensitive materials. Knock-on displacement of light elements (such as oxygen) and mild radiolysis remain possible, but high radiolytic degradation is unlikely at the doses used. While the dose-distribution concepts described above are general, the observed differences between scan strategies in our datasets likely arise from instrument-related factors, such as coil hysteresis, flyback delays, and probe settling dynamics, rather than from active suppression of chemical bonds or volatilization.

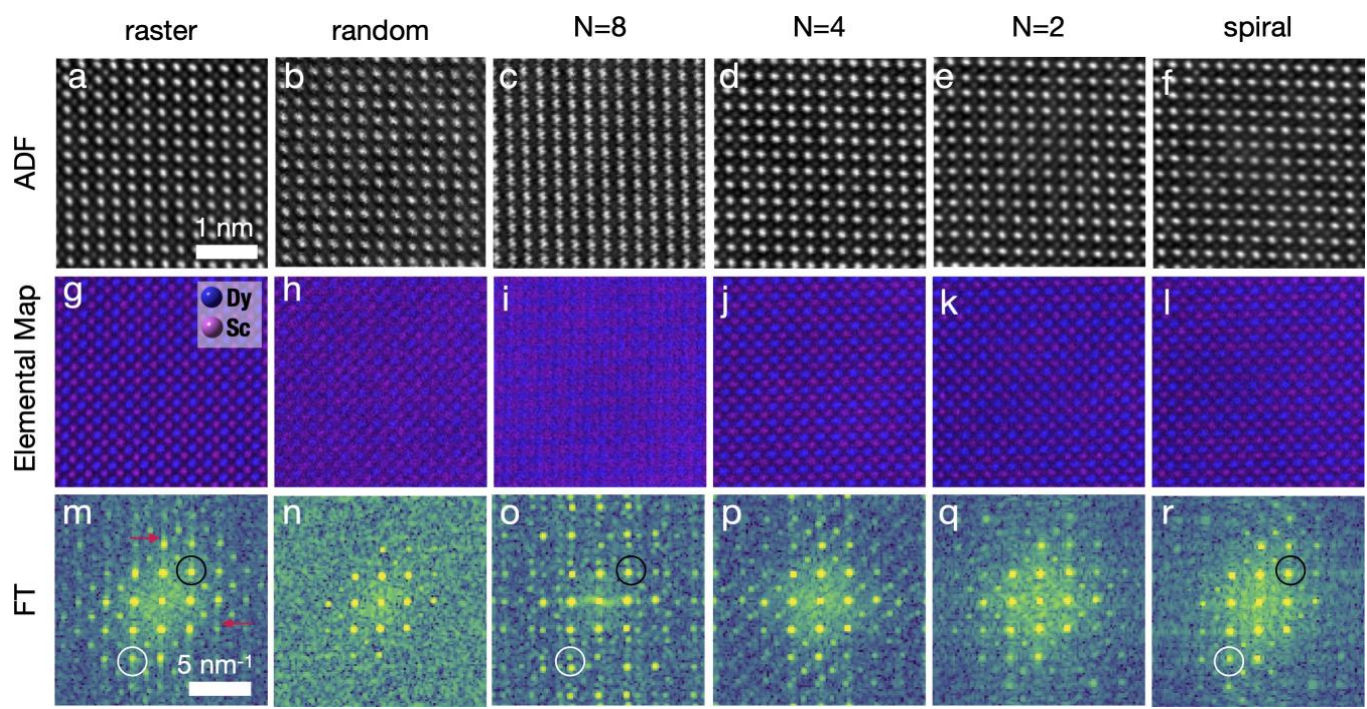

Fig. 3: (a-f) ADF-STEM images of $DyScO_3$ acquired using different scan trajectories (raster, random, N=8, N=4, N=2 sequential, and spiral), together with their corresponding EELS elemental maps (Dy in magenta and Sc in blue) from g-l, and FFTs of the ADF images (m-r). Elemental maps highlight trajectory-dependent variations in signal continuity and distribution. FFTs reveal characteristic artifact signatures: raster scans show vertical streaks from flyback motion, N = 8 sequential scans produce replicated Bragg peaks from large probe jumps, and spiral scans suppress directional streaking, yielding cleaner reciprocal-space features. Characteristic Bragg peaks for raster, N = 8, and spiral scans are circled in black and white to illustrate the distinct artifact patterns introduced by each trajectory.

## Scan-Induced Artifacts in ADF and EELS

Programmable scan trajectories influence both image fidelity and spectroscopic reliability. Mechanical and electronic factors, such as coil hysteresis, probe-settling delays, and flyback motion, introduce position-dependent errors. Temporal variations in dose delivery affect local signal statistics and damage kinetics. Together, these effects manifest as scan-specific artifacts in both ADF and EELS datasets.

High-$N$ sequential scans represent the "worst case" scenario. The large probe jumps produce transient positional errors: the commanded and actual probe positions deviate during rapid transitions, creating abrupt discontinuities between adjacent atomic columns in real space. In reciprocal space, these same errors appear as replicated or shifted Bragg reflections, visible as "false" periodic peaks in the FFTs of ADF images (Figure 3(o)). Because EELS maps are assembled from the same positional metadata, these artifacts propagate directly into elemental maps, degrading both contrast uniformity and quantitative reliability [15, 17].

Scan-induced artifacts (such as those caused by high-$N$ probe jumps or lag) and aliasing both manifest as additional or misplaced features in the FFT, but their physical origins differ. Aliasing arises whenever the spatial sampling rate is insufficient to capture the highest spatial frequencies present in the sample. In high-$N$ sequential scans (e.g., $N = 8$), the large, rapid probe jumps effectively reduce the local sampling density by skipping positions, causing undersampling relative to the Nyquist criterion. This undersampling folds high-frequency information back into lower frequencies in the FFT, producing artificial peaks (classic aliasing artifacts). Meanwhile, scan artifacts from probe lag, coil hysteresis, or synchronization errors tend to create more irregular distortions or streaking that do not strictly follow aliasing folding rules. In practice, these effects often overlap, since scan artifacts that degrade spatial sampling density can directly induce aliasing.

Mitigation of these artifacts requires ensuring that the maximum distance between adjacent sampled points satisfies the Nyquist criterion for the desired spatial resolution, which may involve reducing step size or increasing probe positions, especially in non-raster trajectories with large jumps. While minimizing dwell time can reduce distortion from lag, it does not prevent aliasing caused by insufficient spatial sampling. Therefore, careful scan design and optimized probe positioning are essential for programmable STEM strategies aiming to suppress both scan artifacts and false features induced by aliasing.

Raster scanning is mechanically simpler but suffers from directionally biased artifacts. Its repetitive line-by-line motion introduces flyback and probe-lag effects, which are visible as faint vertical streaks in the FFTs of ADF images (red arrows in

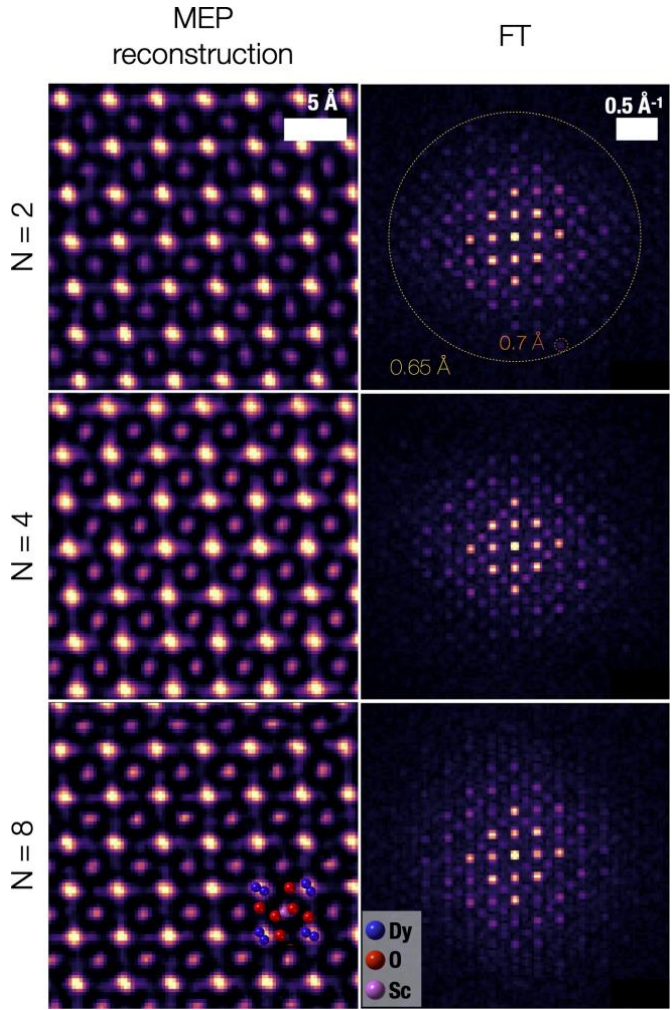

Fig. 4: Ptychographic phase reconstructions of DyScO$_3$ obtained using mixed-mode multislice algorithms (2 probe modes, 18 slices, 2 nm thickness, 20 nm defocus). Comparisons are shown for N-sequential scan patterns (N = 2, 4, 8). FFTs of the reconstructions include a reference circle at 0.65 Å, with visible Bragg peaks extending to ∼0.7 Å. High-resolution information remains consistent despite increasing scan-sequence complexity.

Figure 3(m)). These streaks represent subtle but systematic spatial frequency distortions that do not self-average, even at relatively slow scan speeds.

Spiral scanning provides a favorable trade-off. By removing the repetitive flyback motion, spiral trajectories suppress the directionally biased streaks observed in raster FFTs, producing cleaner, isotropic reciprocal-space data (Figure 3(r)). The smoother coil trajectories reduce transient probe-position errors and yield structurally more reliable ADF and EELS maps. However, spiral scans can introduce a mild radial dose gradient, with slightly higher cumulative exposure near the start or end of the trajectory, depending on whether the scan proceeds inward or outward. This gradient may locally accelerate dose-dependent processes such as radiolysis or transient desorption. In practice, the benefits of reduced directional artifacts and improved spatial frequency fidelity outweigh this minor dose variation, resulting in datasets that are visually cleaner and quantitatively more robust.

Among the modalities we evaluate, EELS is uniquely sensitive to both categories of artifact. Its quantification depends on precise spatial registration and consistent, locally averaged spectral statistics. Temporal non-uniformity in dose—such as inward spirals concentrating electrons near the center or sequential sub-blocks repeatedly re-exposing certain regions—can perturb spectral ratios, alter peak intensities, and bias elemental maps. Even subtle variations in dose delivery timing can compromise quantitative interpretation, amplifying the patchy elemental contrast observed in Dy and Sc maps for $N$ = 8 and some spiral trajectories, even when ptychography remains high-resolution. ADF imaging is also affected, but the consequences are primarily visual (contrast modulation and streaking) rather than quantitative.

Together, these observations demonstrate that while programmable scan strategies offer powerful control over dose distribution and artifact suppression, their practical impact depends on the modality: EELS demands both spatial precision and temporally uniform illumination, whereas ADF tolerates modest non-uniformities, and ptychography remains robust as long as probe overlap and reconstructable positional metadata are preserved.

### Ptychography Tolerates Large Pixel Jumps

Ptychographic reconstruction differs from conventional STEM real-space imaging approaches in that it does not depend on a single pixel-by-pixel intensity recorded on a bright or dark field detector. Using redundant and overlapping diffraction measurements, iterative optimization algorithms can estimate and correct moderate scan path errors [4, 6, 7, 38]. As long as probe overlap, reciprocal-space coverage, and diffraction SNR remain sufficient, high-resolution phase information can be recovered even from non-ideal or irregular trajectories.

Our DyScO$_3$ datasets illustrate this robustness: N = 2, 4, and 8 sequential scans produced phase reconstructions with nearly identical resolution (Figure 4), despite increasingly distorted ADF and EELS counterparts. This resilience reflects the fundamental advantage of ptychography for programmable scanning workflows: trajectory flexibility without sacrificing structural fidelity when there is sufficient dose and overlap between probe positions. However, for lower-dose scans, it may be more challenging to converge reconstructions.

Our findings are consistent with this picture. EELS and ptychography respond differently to non-uniform probe motion. EELS maps show strong sensitivity to temporal and spatial dose irregularities, particularly in high-$N$ sequential scans where large probe jumps induce local contrast discontinuities. In contrast, ptychographic reconstructions remain robust as long as probe overlap and accurate (or reconstructable) positional metadata are preserved [4, 6, 7].

### Cryogenic Spiral Scanning in Free-Standing BaTiO$_3$: Resolving Oxygen Octahedral Shifts

BaTiO$_3$ undergoes a series of temperature-dependent structural phase transitions, including cubic, tetragonal, orthorhombic, and rhombohedral symmetries, driven by displacements of Ti atoms within the TiO$_6$ octahedra. This gives rise to spontaneous polarization and ferroelectric behavior. At low temperature, reductions in crystallographic symmetry are expressed through subtle atomic shifts, particularly of the oxygen sublattice, which define the octahedral tilt pattern and associated dipole moment.

Our cryogenic experiments demonstrate that programmable spiral scanning provides an advantage in resolving these delicate structural changes. Ptychographic reconstructions from conventional raster scans resolved Ba and Ti columns and only faint oxygen contrast (Figure 5(b)). In contrast, outward spiral scanning not only improved oxygen column visibility but also revealed a slight displacement of oxygen atoms consistent with an octahedral shift (Figure 5(c)). This shift, coupled with subtle Ti displacements within the $TiO_6$ framework, implies the presence of a low-temperature ferroelectric phase rather than the centrosymmetric cubic phase observed at room temperature. However, the depth resolution of ptychographic reconstructions is limited, and these measurements are sensitive only to projected 2D displacements. We can therefore confirm the presence of in-plane oxygen shifts, but a full three-dimensional displacement pattern cannot be resolved within this dataset.

The detected displacement pattern is consistent with the onset of a rhombohedral-like phase, as described in recent reports [39]. While 2D projections cannot fully resolve three-dimensional displacement vectors, the observed symmetry breaking and oxygen octahedral shift are compatible with this interpretation. Importantly, programmable spiral scanning was essential for achieving the required signal-to-noise ratio and for suppressing scan-induced artifacts that would otherwise obscure these subtle features.

Looking forward, integration of multislice ptychography with programmable scanning could enable full 3D mapping of octahedral tilts and cation displacements, providing a direct pathway for quantifying ferroelectric polarization at the atomic scale under cryogenic conditions. Such capability would represent a critical advance for studying other phase transitions, polarization dynamics, and strain-engineered ferroelectrics with minimal beam-induced damage.

### Scan Strategy Robustness Across Modalities

Taken together, our results suggest that programmable scanning is broadly compatible with high-resolution STEM, but its optimal implementation depends on the target modality:

- **Ptychography**: tolerant of large probe jumps and trajectory irregularities, enabling flexible dose-distribution schemes without resolution loss.
- **ADF and EELS mapping**: require careful design to avoid spatial misregistration, temporal dose gradients, and coil-settling artifacts that propagate directly into image and spectral contrast.

Across both DSO and BTO datasets, programmable scan patterns, including spiral and N-sequential trajectories, produced high-quality ptychographic reconstructions without introducing structural distortions or signal degradation. Despite significant changes in the probe path, atomic resolution was maintained. FFT magnitudes of the reconstructions confirmed that the highest resolvable Bragg peaks remained essentially unchanged, even for N = 8. In contrast, vBF images from the same datasets revealed clear scan-pattern artifacts, particularly at N = 8, where large probe jumps and sparse sampling produced banding, pixelation, and non-uniform contrast.

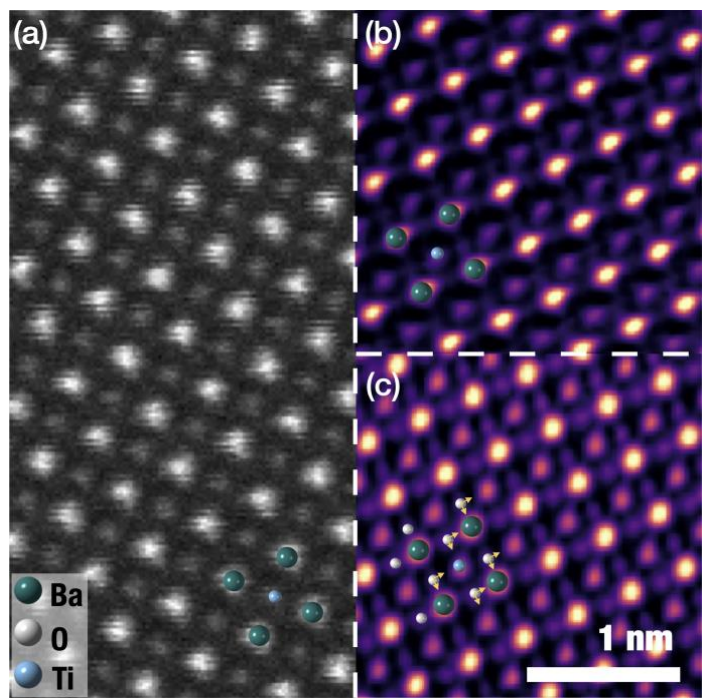

Fig. 5: Comparison of HAADF (a) and ptychographic reconstructions (b,c) of a free-standing $BaTiO_3$ thin film acquired at cryogenic temperature. (b)Raster scan. (c)Outward spiral scan. Oxygen columns are resolved in the spiral-scan ptychographic image, and no significant scan-induced artifacts are visible, demonstrating the suitability of programmable scanning for dose-sensitive cryogenic experiments.

Future work could strengthen these comparisons by including dose quantification maps, controlled probe current variation, and real-time drift correction across programmable trajectories. Nevertheless, our findings demonstrate that programmable scanning is a reliable approach for high-resolution 4D-STEM, EELS, and ptychography, even under non-standard trajectories.

## Conclusion

Programmable electron-beam scanning provides an alternative tool to control both the spatial and temporal distribution of dose in scanning transmission electron microscopy. By systematically comparing raster, spiral, and N-sequential trajectories across atomic-resolution EELS, ADF, and ptychographic datasets, we establish practical guidelines for choosing scan strategies in dose-sensitive workflows.

Our findings highlight three key insights:

1. **Ptychography tolerates scan irregularities.** Iterative phase retrieval consistently recovers high-resolution structural information even from scan patterns with large probe jumps and non-uniform sampling, enabling dose redistribution without sacrificing spatial resolution.
2. **EELS and ADF imaging remain sensitive to trajectory.** Large positional jumps, flyback delays, and non-uniform temporal dose distributions introduce measurable contrast discontinuities and replicated Bragg features that propagate directly into spectroscopic quantification.

3. **Spiral scanning offers a practical compromise.** By eliminating flyback-induced line artifacts while maintaining smooth coil motion, spiral trajectories yield cleaner Fourier space representations and more uniform real space fidelity than raster, with only minor radial dose effects. This proved particularly advantageous for low-dose or cryogenic applications.

These results provide a framework for matching scan design to experimental priorities. Trajectory flexibility can be paired with ptychography for good phase resolution. On the other hand, quantitative EELS and ADF imaging demand careful control of both spatial registration and temporal dose uniformity.

More broadly, programmable scanning is emerging as a central tool for next-generation high-resolution microscopy. As hardware and software mature, including the integration of active feedback, optimized subscan scheduling, and real-time dose control, these approaches will enable atomic-scale characterization in regimes previously inaccessible to electron microscopy. Although this study focused on robust oxide perovskites, the same principles open opportunities for testing programmable scan strategies on truly beam sensitive systems such as polymers, molecular crystals, and biological macromolecules, offering a path toward reliable structural imaging and spectroscopy in materials long considered beyond the practical reach of atomic-resolution analysis.

## Materials and Methods

### Samples

Two model perovskite systems were used. $DyScO_3$ single crystals served as a robust test platform for benchmarking programmable scan performance under atomic-resolution conditions. Free-standing $BaTiO_3$ thin films were used to evaluate low-dose, cryogenic imaging of lighter elements in a mechanically unsupported geometry.

### STEM Data Acquisition

All STEM datasets were acquired on a probe corrected Thermo Fisher Spectra 300 microscope operating at 300 kV, equipped with a Gatan Continuum energy filter and K3 direct electron detector. The K3 detector was operated in electron counting mode. The data for both EELS and 4D-STEM was collected in EFTEM mode at different camera lengths as detailed below. The convergence semi-angle was approximately 30 mrad. The probe current for each mode is also detailed below.

Programmable scan patterns, including conventional raster, spiral (inward and outward), and N-sequential scans (N = 2, 4, 8), were scripted in Gatan Digital Micrograph and used for data collection. Datasets were collected on $DyScO_3$ single crystals and free-standing $BaTiO_3$ thin films. The $DyScO_3$ samples served as a reference for evaluating spatial resolution and reconstruction fidelity under different scan trajectories, while $BaTiO_3$ datasets were acquired under cryogenic conditions (100 K) to test the ability of programmable scans to resolve light elements under beam-sensitive conditions.

### EELS Data Acquisition

The probe current during 4D-STEM acquisitions was maintained at ∼40 pA to preserve low-dose imaging conditions.

STEM EELS datasets were operated in energy-filtered transmission electron microscopy (EFTEM) mode with a 145 mm camera length and a dispersion of 0.9 eV/channel. A subscan mode of 4 × 4 was used, with a scan area of 128 × 128 pixels and a step size of 0.4 nm.

Each dataset comprised 15 passes over the same scan area, with drift correction applied between passes to maintain spatial registration. Under these conditions, the cumulative electron dose for a complete dataset was approximately $1.6 \times 10^7 \ e^-/Å^2$, consistent with typical atomic-resolution spectrum imaging. All programmable scan modes (raster, spiral, and N-sequential) were applied under identical acquisition parameters. Dual EELS acquisition provided simultaneous access to both the low-loss and high-loss regions of the spectrum. The high-loss spectra, after alignment and denoising, are shown in Supplementary Figure S2, where two characteristic peaks are resolved between 1260 and 1400 eV.

For each EELS acquisition, an ADF signal was recorded in parallel. These ADF images were used to assess scan-pattern–dependent artifacts. Hanning filtering, quartile-based thresholding, and cropping were applied to enhance trajectory-specific features. After processing, replicated Bragg reflections and directional streaking became visible, enabling direct comparison of the structural and spectroscopic consequences of each scan strategy.

### 4D-STEM Data Acquisition

The probe current during 4D-STEM acquisitions was maintained at ∼5 pA to preserve low-dose imaging conditions.

The scan areas typically consisted of 64 × 64 probe positions, with diffraction patterns recorded at each point. The probe step size was 0.7 Å, corresponding to a field of view of approximately 4.5 nm. A pixel dwell time of 3.3 ms was used for most acquisitions. The defocus was varied between 10 and 20 nm, optimized manually for each dataset. Variations in sample thickness and defocus were minimized, but not eliminated, across the different scan pattern datasets.

For a representative ptychographic dataset (64 × 64 scan, 0.7 Å step size, 3.3 ms dwell, 5 pA probe current), the electron dose was approximately $2.1 \times 10^5 \ e^-/Å^2$. This corresponds to ∼ $1.0 \times 10^5$ electrons per probe position and a total of ∼ $4.2 \times 10^8$ electrons distributed over a 4.5 nm field of view. These values are consistent with dose regimes reported for low-dose 4D-STEM and EELS workflows [8], enabling direct comparison of dose distribution among programmable scan strategies.

### Ptychographic Reconstructions

Ptychographic reconstruction was performed using the py4DSTEM software package, employing a multi-slice mixed-state algorithm with 10–20 slices (2 nm per slice) [6, 7]. For the $DyScO_3$, two probe modes were used, while for the cryogenic $BaTiO_3$ reconstructions six probe modes were required to accurately describe the experimental probe. The reconstructed probe states for both systems are shown in Supplementary Figures S3 and S4, respectively. All reconstructions were performed with consistent settings to facilitate comparison. Reconstructions were evaluated using phase image analysis and Fourier transforms (FFTs) to assess the highest resolvable Bragg peaks.

Acknowledgments

This work was made possible by the EPSRC Cryo-Enabled Multi-microscopy for Nanoscale Analysis in the Engineering and Physical Sciences EP/V007661/1. M.P., Y.L. and M.S.C. acknowledge funding from the Royal Society Tata University Research Fellowship (URF\R1\201318) and Royal Society Enhancement Award RF\ERE\210200EM1. G.T. EPSRC Centre for Doctoral Training in the Advanced Characterisation of Materials (CDTACM)(EP/S023259/1) for funding their Ph.D. studentships and G.T. acknowledges Cameca Ltd. for co-funding their PhD. Work at the Molecular Foundry was supported by the Office of Science, Office of Basic Energy Sciences, of the U.S. Department of Energy under Contract No. DE-AC02-05CH11231.

Author Contributions Statement

M.S.C. and M.P. collected the experimental data. M.P., L.S., and M.S.C. processed the EELS data. M.P. processed the ptyhcography data using py4DSTEM with input from S.R., C.O. and M.S.C. All authors contributed to the project planning and writing.

## Supplementary Information

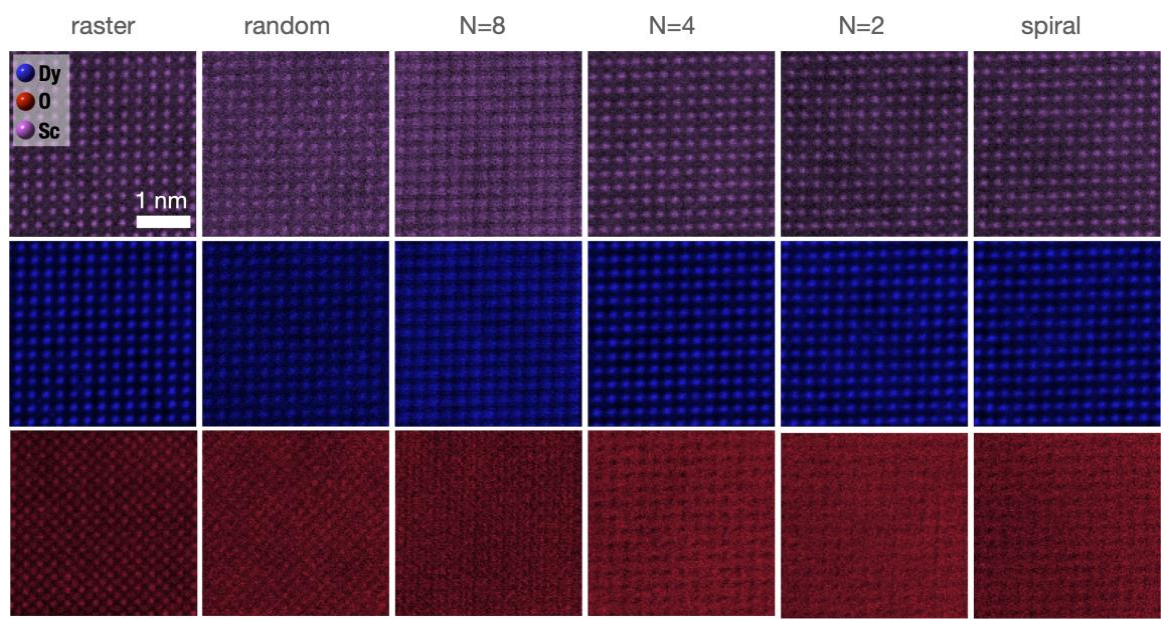

Fig. S1: Elemental EELS maps of Dy (blue), Sc (pink), and O (red) acquired from $DyScO_3$ using different scan trajectories (raster, random, N = 2, N = 4, N = 8 sequential, and spiral).

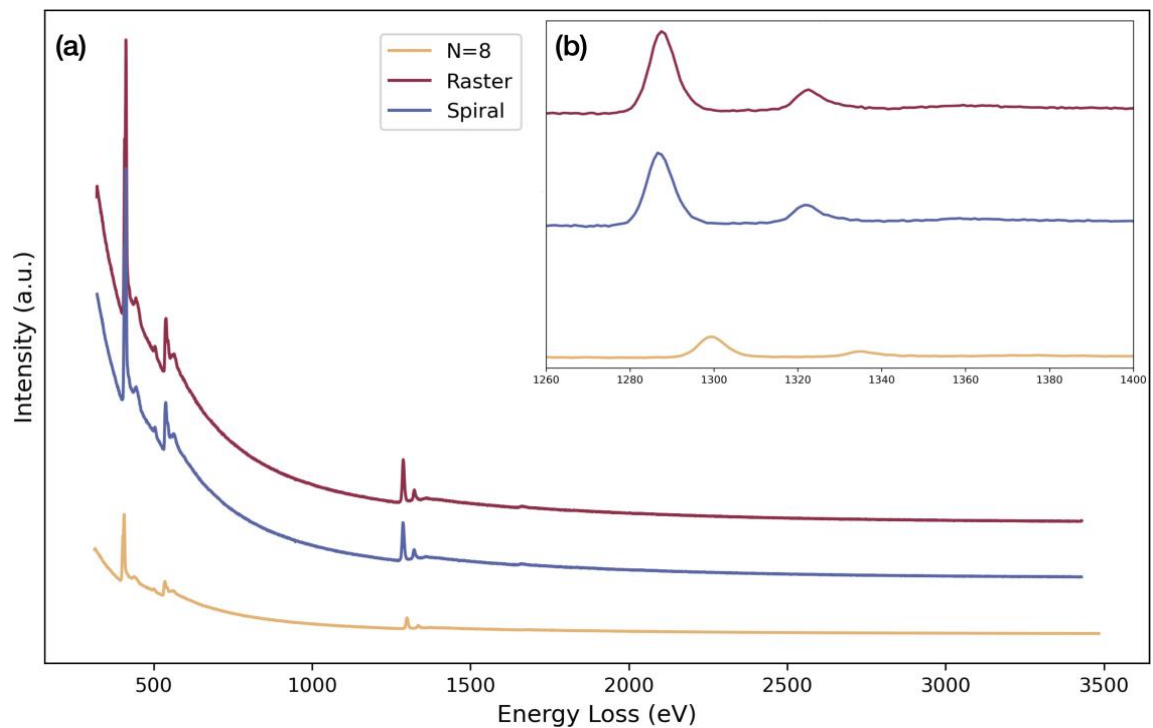

Fig. S2: High-loss EELS spectra from $DyScO_3$ acquired using raster, spiral, and N = 8 sequential scan trajectories. The full spectra (a) are shown together with a magnified region between 1260 and 1400 eV (b), where two peaks are visible.

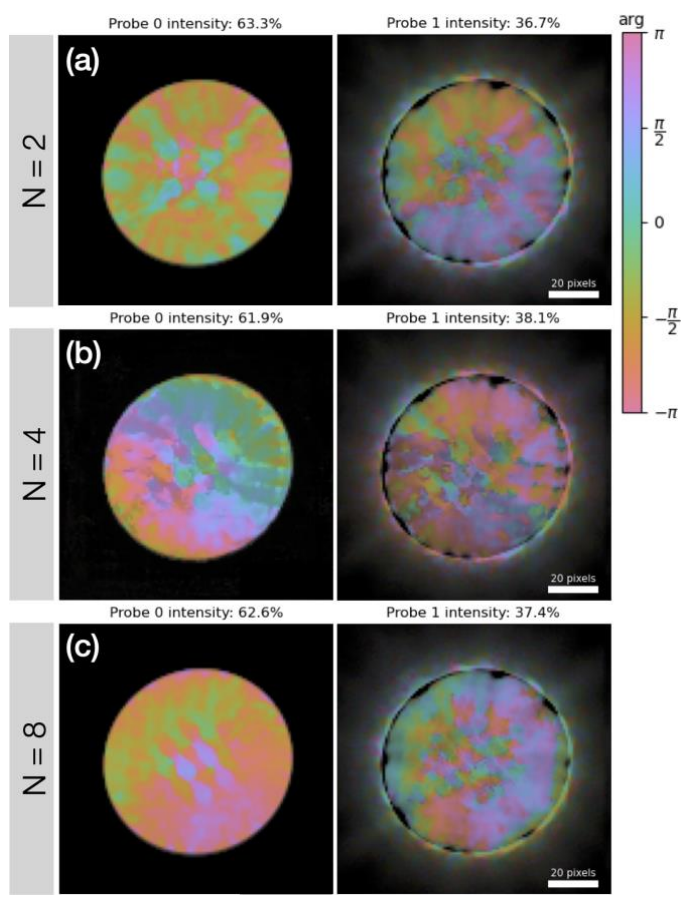

Fig. S3: Reconstructed probe modes from multislice ptychographic analysis of DyScO$_3$. The two dominant probe states are shown for (a) N = 2 (b)N = 4, and (c)N = 8 sequential scan trajectories, illustrating how multiple probe modes are required to accurately describe the experimental electron probe.

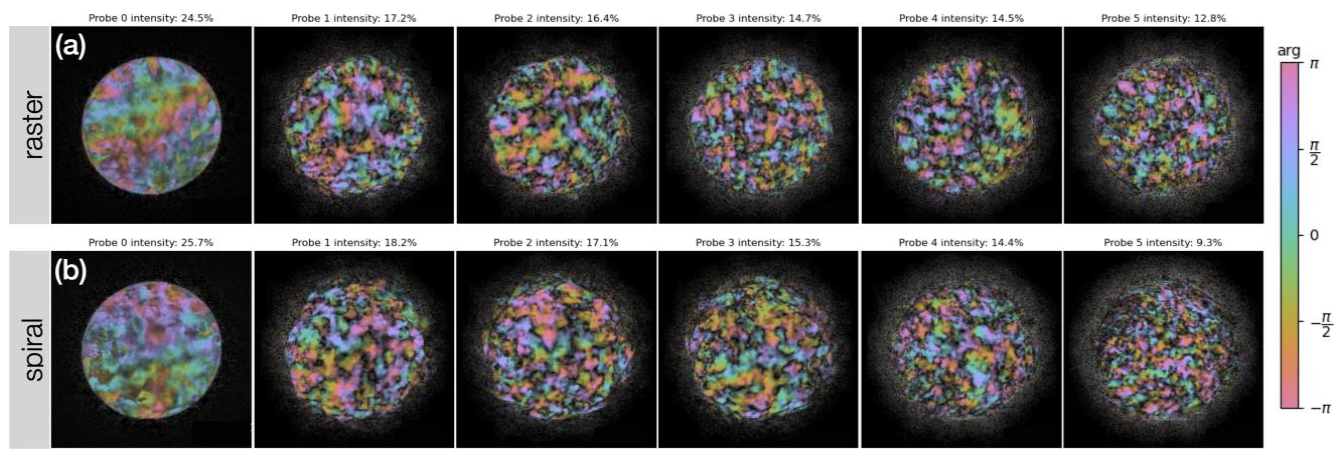

Fig. S4: Reconstructed probe modes from multislice ptychographic analysis of cryogenic BaTiO$_3$ thin films. Six probe states are shown for (a) raster and (b) spiral scan trajectories.